**Surviving Early Career Research and Beyond in the Physics of Life:**

**A concise user guide**

*Mark C Leake[1]*

[1] *Departments of Physics and Biology, University of York, United Kingdom.*

Early Career Researcher (ECR) development is a dynamic challenge that tensions the urge to perform ground-breaking research against an ultimate practical aspiration of establishing an acceptable level of job security. There is no typical career path for an ECR, least of all in the area of the Physics of Life or Biophysics/Biological Physics. Being explicitly interdisciplinary across the physical-life sciences interface presents more opportunities for a multiplicity of career trajectories through different home academic institutions and departments, as well as offering a broader range of alternative future career trajectories in non-academic sectors. That said, there are key common features, such as the transient nature of fixed-term postdoc contracts, the substantial research and domestic challenges that these present, and the often overwhelming pressures of the realities of competition in the job market. In this short article I outline the key challenges to ECRs in the Physics of Life and discuss simple strategies to manage and potentially overcome them.

**Planning**

The one overarching strategy that will best help an ECR to progress in the Physics of Life (PoL) is simply this: *have a plan*. There are multiple aspects of ECR life that are beyond your control; you cannot simply magic up a dream and fully tenured job, nor extend your current postdoc contract *ad libitum*, and the level of influence you can have over the research vision of your supervisor whose project you are currently engaged in can be variable that may become a tension to your pursuing your own independent research aims. But what you can do is at least consider what you ultimately want to do in life – know what your desired destination is, then you can try to work up a framework that may help you in getting to that point from where you currently are. Using something like a workplan or Gantt chart can be enormously helpful in this regard. These do not need to be intensely granular, but they need at least to have some description of a few focused activities, when these are planned to occur, and have a few checkpoints or milestones at around the ~2/3 stage of each activity to allow you to critically gauge how well that activity is going or not, and if not then if you may need to consider pushing forward with an alternative activity as a mitigation plan. One easy way to mitigate against general risk of this nature is to ensure that there is some overlap at all time points in your plan between different activities so that this responsive migration between activities can occur.

One key feature of a plan is that it can, and should, be fluid. Knowing just how fluid to make it is not intuitive however, since the primary function of a plan is to act as a reasonably firm structure that can be followed. But a key time to review your plan is at each milestone checkpoint – use these not only to appraise how well a particular activity is progressing but also use this time to take stock of the global picture, and do not be afraid to adapt the plan as appropriate. This is particularly important in the instance of unforeseen external circumstances influencing the outcome and relevance of your planned research activities.

And consider having more than one plan, to cover different time scales. A long-term plan may be something that can be relatively light on detail but will take you from where you are now, to where you ultimately wish to be, such as a fully tenured academic job that could be several years down the line. A mid-term plan can potentially cover your current ECR post and the broad activities you have planned for this period, which can be both research and non-research based, and a short-term plan can be particularly relevant for ECRs who are engaged in multiple different activities and need to detail these more acutely and over a more aggressive time scale.

A particularly important feature of a PoL ECR plan is to structure in liberal periods for engaging your peers across physical-life sciences interface. This is invaluable, since a driving force for PoL research is ultimately to arrive at some level of truth about the natural biological world, but how you arrive at that truth, and how "true" the truth is when you get there, depends in no small part on where your research journey starts. You need to hone a "bilingual" ability of scientific understanding across the physical-life sciences interface, to be able to at least communicate the same basic ideas between physical sciences and biology. The most rewarding and efficient way to achieve this it to talk to people from across that interface, describe your research to them and get them to describe it back to you in their own words, and swap roles. This resonates with the core aims of Physics of Life research, but will also help you more palpably to, for example, rework your own ECR fellowship proposals to align with different funder remits, if the time is right for you to tackle these.

It sounds like a lot of effort when you are probably pushed for time as is, but a common reason for being pushed for time is the absence of planning and structured prioritizing. So think of a plan not so much as an option, but more as a lifeline.

**Strategy**

Part of your long-term vision as a PoL ECR demands levels of strategic decision making - you need to make tough decisions about balancing competing demands on your time. There are aspects of the core research you are currently engaged in, but there are also other activities related both directly and indirectly to your research that may help you to differing extents to achieve your ultimate goal of job security. You need to be strategic and prioritise publications. A common pitfall of ECRs who wish to progress to ultimately running their own independent research team is to allocate too much time to activities that will take them away from publishing excellent peer-reviewed original research articles. For example, teaching work, excessive attendance at conferences and workshops, acting as an ECR rep on a range of different departmental/institutional/societal committees, and even outreach. These are all valuable activities, but the key is capture balance and appropriate targeting in all of these that will take away your precious time and slow down the process of you publishing excellent research papers. Although metrics that involve these additional activities are helpful for when you apply for your own research funds and/or tenured positions, the reality is that the most substantive, objective and quantifiable way to measure the research ability of an ECR is to look at their publication track record. So, you do need to prioritise getting those papers out.

But then you also need to be strategic in what papers you publish and how many. A single peer-reviewed higher impact and often complexly themed research article can have a significantly more positive influence on ECR fellowship proposals than multiple lower impact publications. By impact, think not of *impact factor* of the journal as such – funders that are likely targets for PoL ECRs to secure their own independent funding, certainly in the UK, now abide by the Declaration of Research Assessment (DORA) (1) which aims to encourage the "responsible use of metrics that align with core academic values and promote consistency and transparency in decision-making", and one feature of this is to steer applicants and reviewers away from using journal impact factor as a mark of how good a particular research article is in that journal. Impact is better measured by the number of times a particular article has been found to be "useful", e.g. the cumulative citations of an article. But such valuable articles will often have multiple and complex strands to them and so may often be cited for a range of different reasons by multiple different communities of researchers. Publishing fewer such high-quality publications is ultimately a better strategy for ECR career progression than multiple shorter, methodological/descriptive pieces, or reviews and book chapters. But such complex research studies often involve multiple collaborators, significant research challenge, and take more time to come to ultimately get published.

What can also make a positive difference is to ensure that you have clearly made a substantive contribution to a paper at the level of being listed as the first or joint-first author. Reviewers for independent funding and/or tenure will often look for clear evidence of an ability to *drive* the research. Having a small handful of cleverly targeted high-quality publications whose research you have clearly led in this way is more convincing evidence for independence and drive than having been a middle author on multiple papers, or indeed a senior/last/corresponding author on a lower impact article. Try not to fall into the trap of thinking that you need to show evidence of being a corresponding author at too early a stage in your research career, focus instead on leading high-quality research from the front, and make sure you finish it off to get it published. This may present challenges if your short-term postdoc contract associated with that research has finished, but there is nothing worse than the sorry tale of the biggest paper that you *almost* published.

It is sensible also to be strategic about identifying gaps in your CV and configuring ways in which to fill them. Such filling could include engaging in some of the non-core ECR activities that I suggested above to rein in on in favour of focusing on publishing great papers.  But again, the key here is balance. Some very focused teaching activities is valuable. Targeted outreach events can be enriching. Directed attendance at conferences and workshops that showcase your own individual excellent research abilities and/or give you specific opportunities to engage potential mentors and champions of your research can be particularly valuable. There may also be learning gaps in your CV, areas of research where your skills and knowledge could be improved. The only way to learn more is to move out of your own research comfort zones. There could for example be opportunities in your current ECR post to learn new techniques through understanding members of your team or from a wider pool of collaborators. You may indeed benefit from identifying new research jobs specifically because they contain some aspects that will force you to develop new skills. The key though is that you need to be proactive, you need to be the one to tell your supervisor, current or future, that being allowed to develop such new skills is of high importance to you.

ECRs often face challenges when it comes to generating "impact", and being able to evidence this on independent fellowship and tenured-post applications. However, they are well placed in being agents of engagement with society, with outreach. These popular science activities in schools, museums, even pubs, can be enriching to both the audience and the speaker. But again though, balance is the key in knowing where to draw the line in not devoting too much of your finite time to any one activity. But participating in focused public engagement can be a fantastic learning experience and does demonstrate clear evidence of the ability to disseminate the broader impact of your research. One of the many beauties of PoL research in terms of outreach is that a lay audience has some intuitive understanding of "living things" in a way that other areas of the physical sciences are arguably harder to discern.

Being strategic in the timing of your own career advancement can also be helpful. For example, in the UK there are cycles of research and teaching assessment in academic institutions known as the Research Excellence Framework (REF) (2) and Teaching Excellence Framework (TEF) (3) respectively. These assessment exercises cycle typicalled over 5+ years, and since the amount of government funds allocated to academic institutions depends about how well they perform in these exercises the run-up to each assessment period traditionally sees many institutions advertising a greater number of targeted tenured posts. Playing the game in this way can verge on being soul-destroying especially if there is not a serendipitous fit aligned to your own domestic life into this *ca.* 5+ year cycle, but being at least forewarned is forearmed, and it does present opportunities to proactively explore potential job opportunities in departments which might not have been on your radar originally, such as considering the value of being hosted by a life sciences department instead of physical sciences, or *vice versa*.

Either way, some general level of strategic targeting of job/fellowship applications is helpful. Taking your time to look for the right targeted fit, as opposed to dispatching a multiple of speculative and unfocused job applications. An obvious strategy to follow here is to look in granular detail at the job description for a given research post. You will find that as a rule there are explicit criteria for each job, some *desirable* others *essential*. If you do not meet all the advertised essential criteria for a given position then an efficient short-listing panel will simply reject your application prior to any interview stage, regardless of how stunning the rest of your application is. Being flexible in regards to where to work can be very helpful, and in many ways moving from your current institution is one clear way to demonstrate independence from your current supervisor. But it is sometimes much easier said than done and dependent upon several domestic factors.

**Money**

Ultimately, how you plan to progress you PoL ECR career depends on the money available to do that, how much and where from. A fantastic idea for your own independent PoL research that is not funded, is just that - a fantastic *idea*. So you need to learn not only how to develop and nurture your own independent research ideas, but also how to follow the money that will enable these research ideas to come to fruition. For UK PoL ECRs, the Engineering and Physical Sciences Research Council (EPSRC) (4) and the Biotechnology and Biological Sciences Research Council (BBSRC) (5) both offer different flavours of ECR fellowships relevant to PoL of typically 3-5 years duration. Other major charitable funders in this area include the Royal Society and Wellcome Trust (for research that is also biomedically relevant). But you need to acquaint yourself with all of the independent research funding schemes relevant to PoL, not just to focus on the long duration and more lucrative fellowships. These include several charities such as the Leverhulme Trust, the 1851 Commission, the Royal Engineering Society, as well as several focused smaller charitable fellowships with specific biomedical themes. These smaller fellowships are often fantastic springboards from which to develop a more compelling proposal for a larger research fellowship to come.

One significant challenge with PoL research, relevant to ECRs, is falling between the gaps of different funder remits. For example, EPSRC might potentially consider your ECR fellowship proposal too biological, BBSRC might triage it as too physical sciences oriented. Arguably, the situation between EPSRC and BBSRC is better now than it was a decade ago, but still there remains a challenge. Here, it is useful to hone a "bilingual" ability across the physical-life sciences interface. EPSRC and BBSRC always recommend discussion your fellowship application in advance with their admin team if there are potential issues of funding remit. However, an even better strategy is talk frequently to researchers from across the physical-life sciences interface about your research ideas. This will ultimately help you to rework proposals to align with different funder remits. A major challenge in PoL research is to work from the often methodological/descriptive structure of physics themed proposals e.g. EPSRC into a more hypothesis–driven framework that is still viewed favourably by funding review panels of BBSRC. Here, you need to try to unpick what the specific key biological questions are to then motivate what research actions you propose to take to address them. This is a skill but it can be learnt from much practice, but in a nutshell it involves training your writing style to be structured more along the lines of "In order to determine X, I will do Y that will demonstrate Z…"

There is also value in demonstrating on your CV the ability to win, or assisting in winning, research funds. As an ECR there are not many opportunities to do this, but they exist. For example, EPSRC and BBSRC grants now allow postdocs who have contributed a significant amount of work into a grant proposal to be named as a "researcher co-investigator" on the award. This is great experience for the postdoc, and also serves as clear evidence to future reviewers of your potential to secure research funds. But even successfully applying for summer studentships, student internships and small pots of internal institutional funding can also serve this ultimate purpose, as well as giving you fantastic experience of grant writing in a relatively time efficient way and, if you then have worked with your supervisor to secure "your" student for a few months you get some invaluable experience at supervising.

**Support**

Finally there is the need to secure support from others for whatever you plan to do in your future PoL ECR career. It is can be enormously helpful to find independent mentors and/or champions or your research ideas, either through your current institution, or through strategically targeting leaders in your field of researchers at conferences or workshops to sell your ideas to them. The COVID pandemic has proved particularly challenging for ECRs in that regard in limiting networking at in-person meetings.

However, one additional resource for engaging with potential mentors and champions is to look toward appropriate networks and learned societies in the area of PoL, the three most relevant in the UK being the Biological Physics Group (BPG) (6) of the Institute of Physics, the British Biophysical Society (BBS) (7) and the Physics of Life network (PoLNET) (8). Signing up as members for all of these bodies can be an enormous investment in many ways for your ECR career progression, giving you access to a range of focused meetings and also a community that wants to really support you.

It can seem dreadfully bleak at times as a PoL ECR, especially when you have less than 1 year remaining on your fixed-term postdoc contract, no concrete offers elsewhere, but a head brimming full of brilliant but perhaps slightly raw research ideas. But there is hope: *there is a wonderful community of PoL researchers in the UK who are genuinely here to support you, you just need to make the effort to reach out to them*.

*Mark Leake is the Chair of Biological Physics at the University of York and Coordinator of its Physics of Life Group. He also serves on the management committee of the British Biophysical Society, and is Chair of the Biological Physics Group of the IoP and of the Physics of Life UK network PoLNET.*